# Thermalisation of a two-dimensional photonic gas in a 'white-wall' photon box

Jan Klaers, Frank Vewinger & Martin Weitz

*Institut für Angewandte Physik, Universität Bonn, Wegelerstr. 8, 53115 Bonn, Germany*

**Bose-Einstein condensation[1] (BEC), the macroscopic accumulation of bosonic particles in the energetic ground state below a critical temperature, has been demonstrated in several physical systems[2-8]. The perhaps best known example of a bosonic gas, blackbody radiation[9], however exhibits no Bose-Einstein condensation at low temperatures[10]. Instead of collectively occupying the lowest energy mode, the photons disappear in the cavity walls when the temperature is lowered - corresponding to a vanishing chemical potential. Here we report on evidence for a thermalised two-dimensional photon gas with freely adjustable chemical potential. Our experiment is based on a dye filled optical microresonator, acting as a 'white-wall' box for photons. Thermalisation is achieved in a photon number-conserving way by photon scattering off the dye-molecules, and the cavity mirrors both provide an effective photon mass and a confining potential - key prerequisites for the Bose-Einstein condensation of photons. As a striking example for the unusual system properties, we demonstrate a yet unobserved light concentration effect into the centre of the confining potential, an effect**

with prospects for increasing the efficiency of diffuse solar light collection[11].

Following the achievement of atomic Bose-Einstein condensation[2-4], we have witnessed interest in light sources where a macroscopically populated photon mode is not the consequence of laser-like amplification, but rather due to a thermal equilibrium phase transition. Work in this direction includes the proposal of a superfluid phase transition of photons in a nonlinear cavity[12-14] and, albeit in the strong coupling regime, the demonstration of a quasiequilibrium phase transition of exciton-polariton quasiparticles to condensates 'half matter, half light'[6-8]. In the weak coupling regime (as in our case), optical cavities have been used to achieve a modified spontaneous emission of atoms and molecules[15-17].

The main idea of our experiment is to study thermalisation of a photon gas, to a heat bath near room temperature (dye molecules), in a system with reduced spatial dimensionality and energy spectrum restricted to values far above the thermal energy. The photons are trapped in a curved-mirror optical microcavity, and repeatedly scatter off dye molecules. The longitudinal confinement (along the cavity axis) introduces a large frequency spacing between adjacent longitudinal modes and modifies spontaneous emission coupling such that basically only photons of longitudinal mode number q=7 (see Fig.1a) are observed to populate the cavity. By this, an effective low frequency cutoff $\omega_{cutoff}$ (the eigenfrequency for the corresponding $TEM_{00}$ transverse mode) is introduced, with $\hbar\omega_{cutoff} \sim 2.1 eV$, much larger than the thermal energy $k_B T$ (~1/40eV at room temperature). The two remaining transverse modal degrees of freedom of light thermalise to the (internal rovibrational) temperature of the dye solution, and the photon frequencies will be distributed by an amount ~$k_B T/\hbar$ above the cavity cutoff. We expect significant population for the TEM$_{nm}$ modes with high transversal quantum number n and m (of high eigenfrequency) at large temperature, while the population concentrates to the lower transverse modes when the system is cold. Equilibrium is reached as photons are absorbed and emitted by dye molecules many times, with the interplay between fluorescence and absorption leading to a thermal population of cavity modes, making the photon gas equilibrate at the temperature of the dye solution (see Methods). During the course of thermalisation (in an idealized experiment) only transverse mode quantum numbers are varied, and - different than in experiments observing blackbody radiation[9,18] - no photons are destroyed or created in average. This becomes clear by noting that the dye molecules have quantized electronic excitation levels with transition energy near or above the cutoff, and thermal excitation is suppressed by a factor of order $\exp(-\hbar\omega_{cutoff}/k_B T)$ ~$10^{-36}$.

The thermodynamics of the system can readily be obtained by a mapping onto the two-dimensional ideal Bose gas: For paraxial propagation and fixed longitudinal excitation number, photons can be treated as non-relativistic massive particles with mass $m_{ph}=\hbar\omega_{cutoff}/c^2$ (see Methods), moving in the transverse plane of the resonator[12]. Due to the curvature of the mirrors the photons are moreover under harmonic confinement (Methods). When the photons are in thermal equilibrium and their average number is conserved, we expect that this trapped two-dimensional ideal Bose gas will undergo a phase transition to a Bose-Einstein condensate at sufficiently low temperature and high density[19,20]. A notable further consequence of the thermalisation of the trapped photon gas is an accumulation in the trap center, where the photon trapping potential is minimum. This behavior is evident from the analogy to a trapped gas of material particles.

Our experimental setup shown in Fig.1b is based on an optical resonator consisting of two high-reflecting curved optical mirrors spaced 3½ optical wavelengths apart, filled with dye solution (see Methods). The short distance between the cavity mirrors yields a free spectral range (~$7 \cdot 10^{13}$Hz) much larger than the thermal energy $k_B T$ in frequency units (~$6 \cdot 10^{12}$Hz at room temperature), and comparable to the spectral width of the dye emission. The mean transversal excitation number (per axis) of the two-dimensional photon gas is $k_B T/\hbar\Omega$~160, where $\Omega/2\pi$ (~$4 \cdot 10^{10}$Hz) denotes the spacing of transversal cavity modes, i.e. the spacing is so small that the transverse motion is quasicontinuous. Other than in a perfect "photon box"[21] we expect losses from coupling to optical modes not confined in the cavity, non-radiative decay and finite finesse. To compensate for the loss rate, the dye is pumped with an external laser beam. In a simple model, the pumping fills a reservoir of electronic excitations in the dye that can exchange particles with the photon gas. Thus the photon gas is seen as an open system, in the sense of a grand-canonical ensemble. The pumping maintains a steady state in which the average photon number $N_{ph}$ will be proportional to the number of electronic excitations $N_{exc}$ and is determined by $N_{ph}/N_{exc} = \tau_{ph}/\tau_{exc}$, where $\tau_{ph}$ and $\tau_{exc}$ denote the lifetimes of photons and electronic excitations, respectively. Despite losses and pumping we expect the photon gas to be well described by an equilibrium distribution when the thermalisation is sufficiently fast, i.e. a photon scatters several times off a molecule before being lost. Our current experiment is carried out at an optical pumping beam intensity of ~10W/cm², an estimated three orders of magnitude below the onset of a photon Bose-Einstein condensate and four orders of magnitude below saturation, so that we do not expect collective photonic effects to be relevant.

Experimental data for the spectral distribution of light transmitted by one of the cavity mirrors is shown in Fig.2a for two different temperatures of the resonator. The connected circles represent experimental data, and the solid lines are theoretical curves for a fully thermalised gas at the corresponding temperature. The expected photon number in the cavity $n_{T,\mu}(u)$ at given transversal energy $u=\varepsilon-\hbar\omega_{cutoff}$ is the product of the energy degeneracy $g(u)=2(u/\hbar\Omega+1)$ (see Fig.1a) and the Bose-Einstein distribution factor:

$$n_{T,\mu}(u) = \frac{g(u)}{\exp\left(\dfrac{u-\mu}{k_B T}\right) - 1}. \tag{1}$$

The chemical potential μ is determined by measuring the output power $P_{out}$=50±5nW, which corresponds to a photon number of $N_{ph}$=60±10 in the cavity, and solving $\sum_u n_{T,\mu}(u) = N_{ph}$ for $\mu$. For the two cases of Fig.2a we obtain $\mu/k_B T$=- 6.76±0.17 (*T*=300K) and *μ/k<sub>B</sub>T*=-7.16±0.17 (*T*=365K) respectively. Because of *μ<<-k<sub>B</sub>T*, the term -1 in the denominator of eq.(1) can be neglected and the distribution becomes Boltzmann-like, with the term $e^{\mu/k_B T}$ acting as a prefactor to the spectral distribution. A Bose-Einstein condensate would be expected for

$$N_{ph} \to N_c = \sum_{u>0} n_{T,\mu=0}(u) = \frac{\pi^2}{3}\left(\frac{k_B T}{\hbar\Omega}\right)^2 \approx 80400$$

(at which $\mu \to 0$). The shapes of the shown experimental spectra are in good agreement with theoretical expectations over a wide spectral range (a visible deviation near 532nm stems from residual pump light). Notably, the enhancement of the short wavelength wing for the higher temperature data by almost an order of magnitude is predicted quite accurately. We interpret these results as one line of evidence for the two-dimensional photon gas to be in thermal equilibrium with the dye solution. Evidence is also obtained from investigating the spatial distribution of light emitted by the resonator. A typical snapshot is shown in Fig.2b, in which a shift from the yellow spectral regime for light near axis towards the green (i.e. higher photon energy) for the radiation emitted off-axis is clearly visible. This is due to the higher energy of cavity resonances with large transversal momentum and correspondingly large mode diameter. From the image in Fig.2b, the spatial intensity distribution shown in Fig.2c was extracted, which is in good agreement with a thermal average over spatial modes. In other words, the thermal distribution of the photon spectrum is also reflected by the spatial distribution.

In further measurements, we have varied the cutoff wavelength $\lambda_{cutoff} = 2\pi c / \omega_{cutoff}$ by piezo tuning of the distance between cavity mirrors. A series of corresponding spectra is shown in Fig.3a. Reasonable agreement with a thermal distribution is obtained for the two spectra shown at the top with cutoff wavelengths near 570nm and 590nm respectively, while the two spectra with longer cutoff wavelength shown in the bottom appear only partly thermalised. We attribute this to weak dye reabsorption in this spectral regime, preventing repeated photon scattering off the dye molecules (also the reflectivity of cavity mirrors lessens at wavelengths above 600nm). This illustrates the importance of both emission and reabsorption of scattered photons for thermalisation, in agreement with expectations.

Finally, thermalisation was investigated by varying the spatial position of the external pumping beam with respect to the centre of the trapping region. Fig.3b shows the position of the maximum of the emitted fluorescence $x_{max}$ versus the transverse position of the pump focus $x_{exc}$ for two different values of the cutoff wavelength. For the data recorded with a cutoff wavelength near 580nm (squares fitted with a green line) the fluorescence essentially freezes to the position of the trapping centre, when the excitation spot is closer than approximately 60μm distance. This effect is not observed when tuning the cutoff wavelength to 620nm (circles fitted with a red line), where weak reabsorption prevents multiple scattering events and thus an efficient thermalisation (see Fig.3a). The demonstrated position locking effect is a direct consequence of the thermalisation, leading to a photon accumulation at the trap centre, where the confining potential imposed by the curved mirrors exhibits a minimum valve. We attribute the observed breakdown of thermalisation for datapoints with

|x$_{exc}$|>60µm to the finite quantum efficiency of the dye (η≅95%)[22], and our mirrors only providing a one-dimensional photonic bandgap, limiting the number of possible photon scattering processes. The relaxation also for initial states further apart from equilibrium could be improved with mirrors providing a full three-dimensional photonic bandgap[23]. Technical applications of the observed directed diffusion effect could include the collection of diffuse solar light to a central spot[14], with prospects for enhancing the optical phase space density. When a Bose-Einstein condensate of photons can be reached, perspectives include coherent UV sources[24].

**Methods**

**Experimental setup**  Our optical resonator consists of two high reflecting dielectric mirrors with spherical curvature (R=1m). One of the mirrors is cut to 1mmx1mm surface size to allow for a distance between mirrors in the micrometer regime ($D_0$≅1.46µm) despite the mirrors curvature. The cavity finesse is ~10$^5$ in the relevant wavelength regime (520nm-590nm). The resonator is filled with a drop of dye (Rhodamine 6G or PDI respectively) solved in an organic solvent and is pumped with a laser beam near 532nm wavelength inclined at 45° to the cavity axis, exploiting a reflectivity minimum.

**Photon dispersion in cavity**  The photon energy as a function of transversal ($k_r$) and longitudinal ($k_z$) wavenumber in the paraxial approximation ($k_z \gg k_r$) is $E = \hbar c \sqrt{k_z^2 + k_r^2} \cong \hbar c (k_z + k_r^2/2k_z)$ with $c$ as speed of light in the medium. Within the resonator, we have $k_z(r) = q\pi/D(r)$ with $D(r) = D_0 - 2\left(R - \sqrt{R^2 - r^2}\right)$ as the mirror separation at distance $r$ from the axis. For $r \ll R$ and a fixed longitudinal mode number $q$ one finds

$$E \cong m_{ph}c^2 + \frac{(\hbar k_r)^2}{2m_{ph}} + \frac{1}{2}m_{ph}\Omega^2 r^2.$$

We arrive at the dispersion of a particle with mass $m_{ph} = \hbar k_z(0)/c = \hbar\omega_{cutoff}/c^2$ subject to a harmonic oscillator potential of trapping frequency $\Omega = c/\sqrt{D_0 R/2}$ in two spatial dimensions.

**Principle of radiation field thermalisation**  For photons absorbed and emitted by dye molecules many times, the state of the light field can be regarded to follow a random walk in configuration space. For a transition between two configurations, $P$ and $Q$ (i.e. sets of mode occupation numbers), induced by the absorption of a photon in mode i and the emission into mode j, the transition rate is expected to follow $R(P \to Q) \propto \alpha_T(\omega_i) f_T(\omega_j)$ with $\alpha_T(\omega)$ and $f_T(\omega)$ as (temperature dependent) relative absorption and fluorescence strength coefficients. This random walk in configuration space will lead to a thermal equilibrium population of photon modes, if the condition of detailed balance[26] is fulfilled:

$$\frac{R(P \to Q)}{R(Q \to P)} = \frac{\alpha_T(\omega_i) f_T(\omega_j)}{\alpha_T(\omega_j) f_T(\omega_i)} = \exp\left(-\frac{\hbar(\omega_j - \omega_i)}{k_B T}\right). \tag{2}$$

Dye spectra are known to typically fulfil (2) within good accuracy, which itself is a consequence of a thermal equilibrium of rovibrational excitations within the ground and excited electronic molecular levels respectively[27-29].

For a simple model of the thermalisation process in the dye, consider a two-level system with energy splitting $\hbar\omega_0$ subject to additional rovibrational sublevel structure of ground and excited states[29]. Absorption $\alpha_T(\omega)$ and fluorescence strength $f_T(\omega)$ at frequency $\omega$ can be expanded by integrating over the contributions from individual energetic sublevels, yielding a ratio

$$\frac{f_T(\omega)}{\alpha_T(\omega)} \propto \frac{\int g'(e')p(e')A(e',\omega)de'}{\int g(e)\exp(-e/k_BT)B(e,\omega)de} ,$$

where $e$ and $e'$ denote the energies of sublevels of ground and excited electronic state respectively, $p(e')$ the population of sublevels in the excited electronic state, $g$ and $g'$ the density of states, and $A$ and $B$ are the Einstein coefficients. The Einstein coefficients at sublevels with energies $e$ and $e'$ matching $\hbar\omega + e = \hbar\omega_0 + e'$, are connected by the A-B relation $g'(e')A(e',\omega)de' = \frac{2\hbar\omega^3}{\pi c^2}g(e)B(e,\omega)de$. When thermalisation in the electronically excited molecular level is sufficiently fast, $p(e') \propto e^{-e'/k_BT} = e^{-(e+\hbar(\omega-\omega_0))/k_BT}$ and one immediately finds $f_T(\omega)/\alpha_T(\omega) \propto \omega^3 e^{-\hbar(\omega-\omega_0)/k_BT}$. With this, the detailed balance condition (2) relating absorption and fluorescence at two different frequencies is fulfilled for $\hbar\omega >> k_BT$.

**Acknowledgements**

We thank F. Schelle for experimental contributions during the early phase of this project. Financial support from the Deutsche Forschungsgemeinschaft within the focused research unit FOR557 is acknowledged.

**Figure captions**

Figure 1: **Cavity mode spectrum and setup. a,** The top graph gives a schematic spectrum of cavity modes. Transverse modes belonging to the manifold of longitudinal mode number q=7 are shown by black lines, those of other longitudinal mode numbers in grey. The energy of an eigenmode in the q=7 manifold is $\varepsilon = \hbar\omega_{cutoff} + u$ with transversal energy $u = \hbar\Omega(n_x + n_y)$, transversal excitation numbers $n_x$, $n_y$ and frequency splitting $\Omega$ between transversal modes[25]. The degeneracy $g(u) = 2(u/\hbar\Omega + 1)$ of a given transversal energy exhibits a linear energy scaling, and the prefactor 2 originates from the two possible polarizations. The bottom graph indicates the (measured) relative absorption coefficient and fluorescence strength of Rhodamine 6G dye versus frequency. **b,** Experimental setup for thermalisation of a two-dimensional photon gas.

Figure 2: **Experimental spectra and intensity distribution**. **a**, The connected dots give measured spectral intensity distributions for temperatures of 300K (top) and 365 K (bottom) of the resonator setup. The solid lines are theoretical spectra based on Bose-Einstein distributed transversal excitations, and for illustration a $T$=300K distribution is also inserted in the bottom graph (dashed line). The measurements shown in this figure were performed with Rhodamine 6G dye solved in ethylene glycol ($c$=5·$10^{-4}$ M). Note that the spectral maximum of blackbody radiation at $T$=300K is at ~10µm wavelength, i.e. far to the red of the shown spectral regime. **b**, Image of the radiation emitted along the cavity axis at room temperature ($T$=300K), showing a shift towards shorter (higher energetic) optical wavelengths for off-axis radiation. **c**, Spatial intensity distribution at $T$=300K (connected circles) along with the theoretical prediction (solid line).

Figure 3: **Break-down of thermalisation**. **a**, Measured spectral intensity distribution for different cavity cutoff wavelengths (connected dots) overlaid with theory curves (solid lines). For the two spectra with longer cutoff wavelengths, the visible deviation from a thermal spectrum is attributed to weak dye reabsorption in this spectral regime (Rhodamine 6G in ethylene glycol, $c$=5·$10^{-4}$ M). **b**, Measured distance of the intensity maximum of the emitted radiation $|x_{max}|$ from the centre versus position of the excitation spot $x_{exc}$. For the data recorded with $\lambda_{cutoff}$≈580nm (squares fitted with a green line) a position locking to the cavity centre is observed, while the fluorescence follows the excitation spot when tuning to weak reabsorption ($\lambda_{cutoff}$≈620nm, circles fitted with a red line) (Perylene-diimide (PDI) solved in acetone, $c$=1g/l).

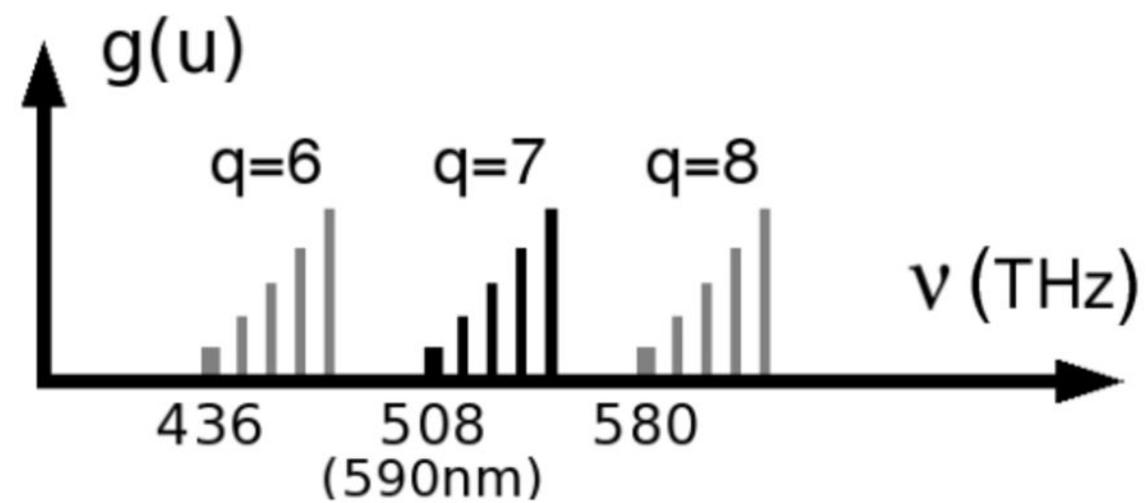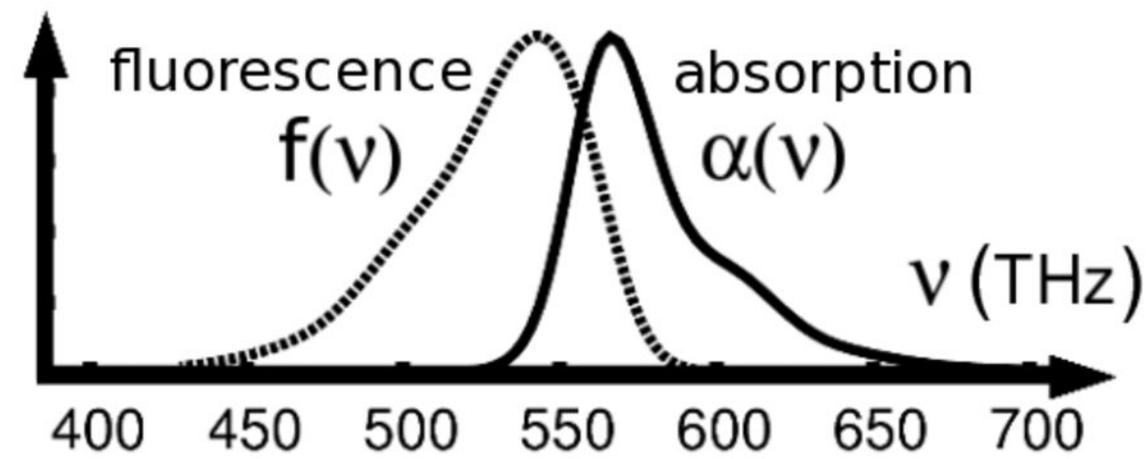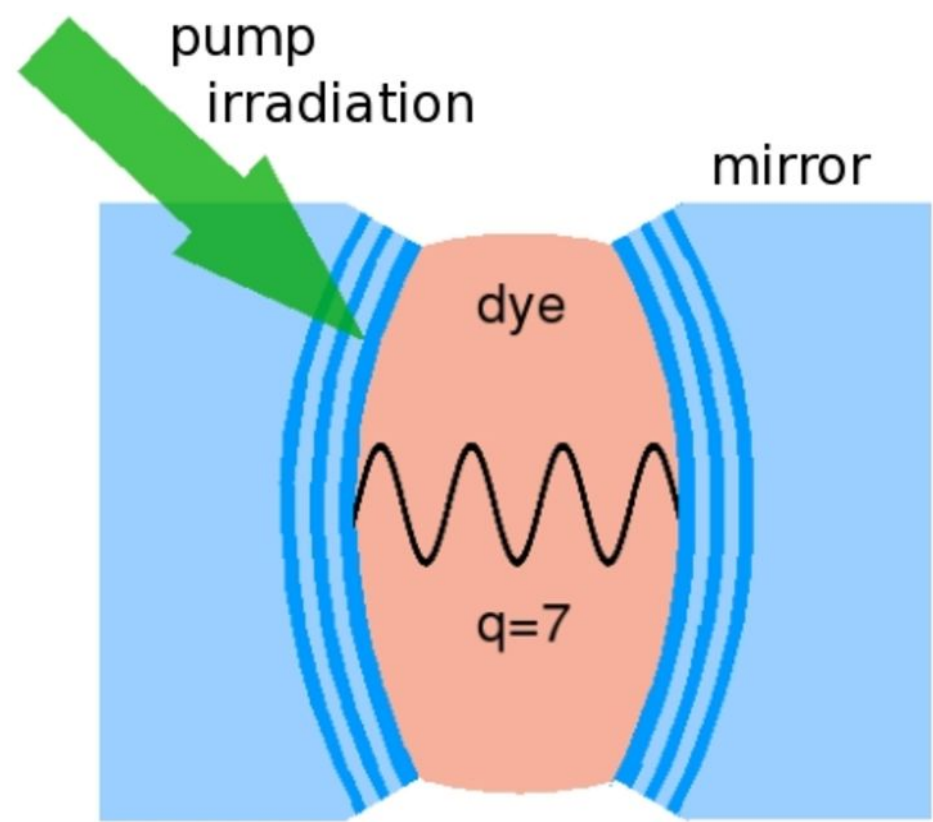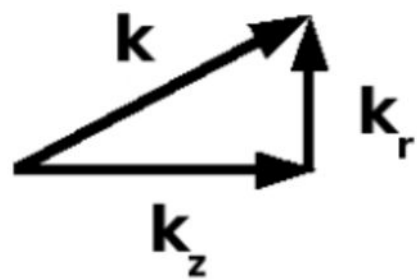

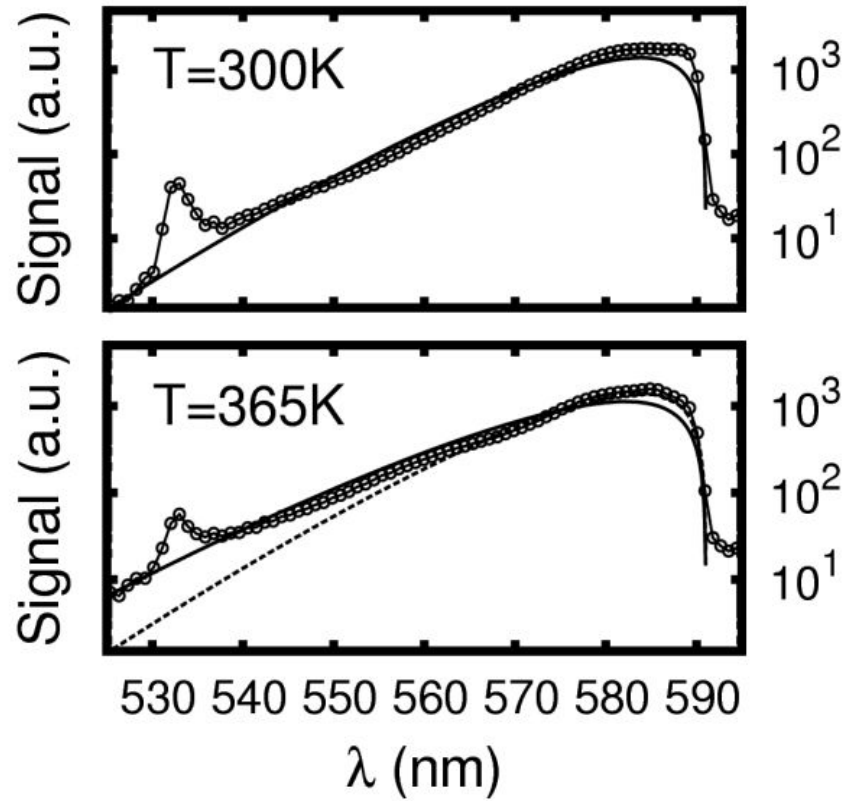
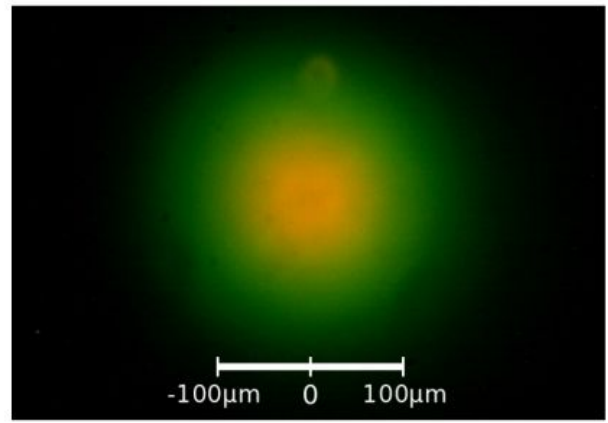
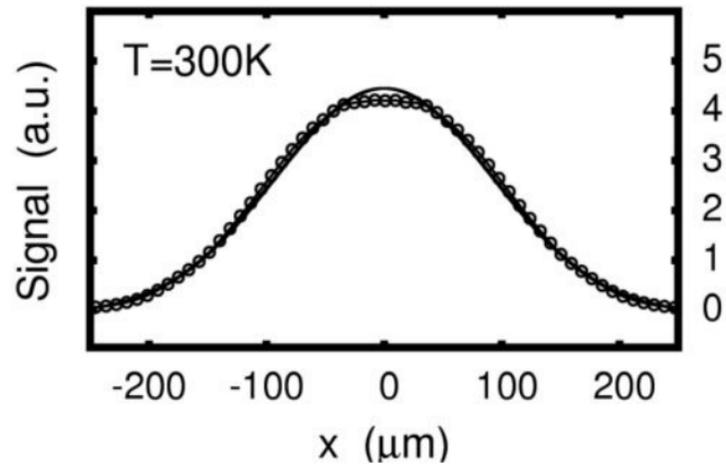

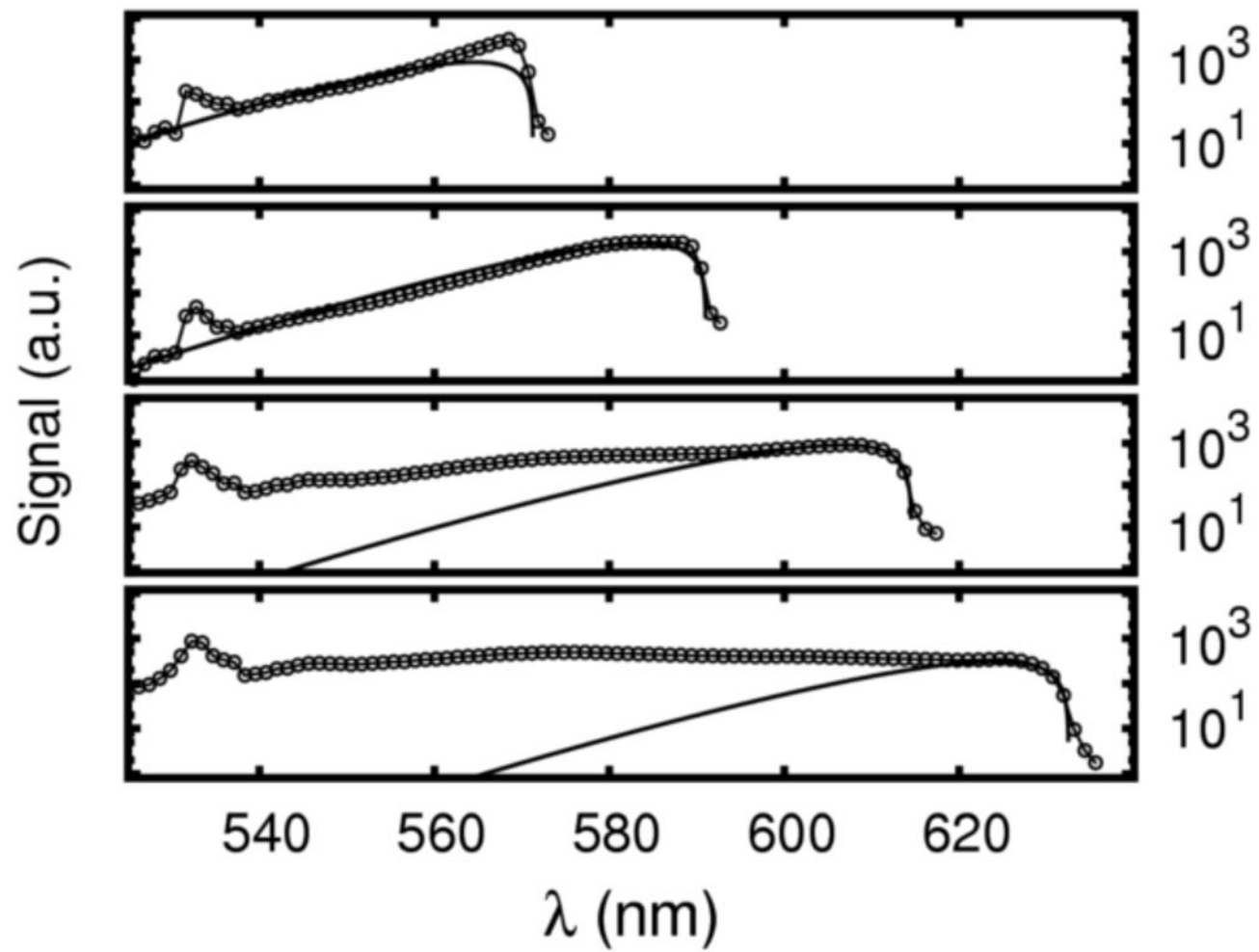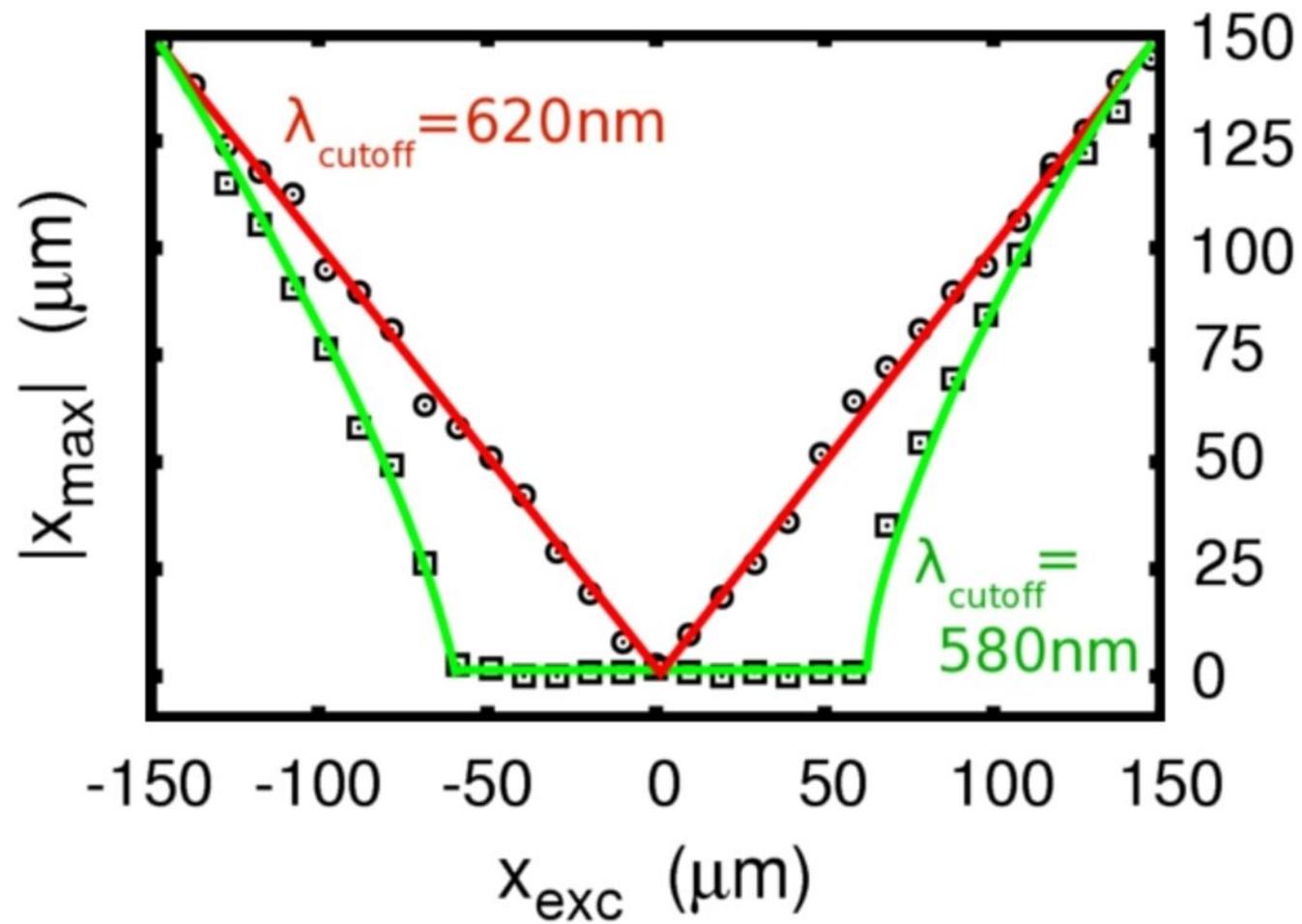